# Electronic and optical properties of FeSe$_2$ polymorphs: solar cell absorber

B.G.Ganga[a], C.Ganeshraj[a], A.Gopal Krishna[b] and P.N.Santhosh[a,*]

[a]Low Temperature Laboratory, Department of Physics, Indian Institute of Technology Madras, Chennai-600036, India

[b]Department of Biotechnology, Indian Institute of Technology Madras, Chennai-600036, India.

Electronic and optical properties of semiconducting FeSe$_2$ polymorphs (marcasite and pyrite) were evaluated using density functional theory with norm-conserving pseudopotential. Marcasite and pyrite of FeSe$_2$ show indirect band gaps of 0.86 eV and 0.67 eV respectively. Absorption coefficient is high over a wide range of optical spectrum, indicating that these materials are exploitable as solar cell absorber. Marcasite FeSe$_2$ is found to be optically anisotropic, whereas pyrite is isotropic. Mulliken Bond Population analysis shows mixed ionic-covalent nature of bonding for both polymorphs. Effective masses of electrons and holes were also calculated, which reinforce the semiconducting nature of FeSe$_2$.

---

[*] Corresponding author.
E-mail address: santhosh@physics.iitm.ac.in

# 1. Introduction

Over past decades, transition metal chalcogenides are experiencing a resurgence of interest due to their excellent optical, electrical, magnetic and transport properties [1-3]. In general, iron sulfide and iron selenide are ideal materials for the fabrication of photovoltaic devices and solar cells due to their suitable band gap (~ 1 eV) and high absorption coefficient (> $10^5$ cm$^{-1}$) [4,5]. FeSe$_2$ thin films grown by electro deposition have optical band gap varies from 0.95 to 1.03 eV with varying film thickness, indicating the importance of this material as an absorber material for thin film solar cell [5]. Moreover, Binxia et al. synthesized flower like FeSe$_2$ which exhibit absorption peak at 1000 nm ($E_g$ = 1.24 eV) and pointed that FeSe$_2$ can be considered as a suitable material for photovoltaic applications [6].

FeSe$_2$ crystallizes both in marcasite orthorhombic structure and pyrite cubic structure and exhibit semiconducting behaviour [7]. Under ambient conditions, FeSe$_2$ adopt marcasite structure while high temperature / high pressure synthesis leads to pyrite type structure [8]. Properties of iron selenides are greatly influenced by composition as well as crystal structure [7]. At room temperature, p-type semiconducting behaviour was observed for stochiometric samples of marcasite FeSe$_2$ and showed an n-type behaviour at high temperature [7]. A better understanding of electronic properties is crucial for technological applications of materials. Electronic and optical properties of FeS$_2$ have been extensively studied but a few experimental investigations are reported for FeSe$_2$. In recent years, a number of theoretical investigations on FeS$_2$ has been published [9-11]. To the best of our knowledge, this is the first attempt to investigate the structural, electronic and optical properties of two polymorphs of FeSe$_2$ using first principles calculations.

Naturally occurring FeSe$_2$ adopt an orthorhombic marcasite structure with space group *Pnnm* (No. 58). The conventional unit cell shown in Fig. 1(a) contains two formula units of FeSe$_2$. Each Fe atom is coordinated by six Se atoms as octahedron and each Se atom is coordinated to three Fe atoms and another Se form a tetrahedron. Two neighbouring octahedra share an edge and form a linear chain parallel to orthorhombic c-axis, as shown in Fig. 1(c) [12].

The high temperature phase of $FeSe_2$ has pyrite like structure with *Pa3* space group (No. 205) is shown in Fig. 1(b). The conventional unit cell contains four formula units of $FeSe_2$. Fe atoms are situated at face centered site of cubic lattice. Cations have the same octahedral environment as in marcasite structure but as we can see from Fig. 1(d), adjacent octahedra are interconnected by corner in pyrite $FeSe_2$. Axes of Se dimers are aligned along [111] direction of cube [12].

## 2. Computational methods

The calculations were performed based on density functional theory (DFT) using CASTEP (Cambridge Serial Total Energy Package) code [13]. The electron–electron exchange and correlation effects are described Perdew-Burke-Ernzerhof (PBE) functional in Generalized Gradient Approximation (GGA) [14]. Norm-conserving pseudopotential is used for our calculations, due to the limitations in calculating optical properties related to ultrasoft pseudopotentials [15]. A plane wave basis set of cut off energy 860 eV is used to expand the valence electronic wave functions. Valence orbitals for generating pseudopotentials were $3d^6 4s^2$ for Fe and $4s^2 4p^4$ for Se. The sampling of Brillouin zone was carefully tested and based on these convergence tests, a k-point grid of 4x3x5 for marcasite phase and 4×4×4 for pyrite phase were used. The geometry optimization was performed with the experimentally observed cell parameters and internal co-ordinates of ions [16,17], until the total energy, maximum force, maximum stress and maximum displacement on the system converge to a tolerance value of $10^{-5}$ eV/atom, 0.03 eV/Å, 0.05 GPa and 0.001 Å respectively. We have calculated band structure, density of states and optical properties on geometrically optimized structure.

Optical spectroscopy is an important tool for studying the electronic structure of materials. The optical properties of matter can be described by means of the dielectric function $\varepsilon(q,\omega)$ where q and ω represent momentum and energy respectively. In the present calculations, we have used dipole approximation and the momentum transfer from the initial state to final state is neglected. Considering contributions only from direct interband transitions, the imaginary part of the dielectric function $\varepsilon_2(\omega)$ is calculated by summing all possible transitions between occupied and unoccupied states with fixed k-vector over Brillouin zone. The transition matrix element representing the probability for the transition is given by [18]

$$\varepsilon_2(q \to O_{\hat{u}}, \hbar\omega) = \frac{2\pi e^2}{\Omega \varepsilon_0} \sum_{k,v,c} \left|\left\langle \psi_k^c \left| \hat{u} \cdot r \right| \psi_k^v \right\rangle\right|^2 \delta\left(E_k^c - E_k^v - E\right)$$

Where $e$ is the electronic charge, $k$ represents k points, $\Omega$ is the volume, $\hat{u}$ is the vector defining the polarization of incident electric field, ω is the photon frequency and $c$ and $v$ represent conduction band (CB) and valence band (VB) respectively. The dispersive or real part $\varepsilon_1(\omega)$ can be obtained from $\varepsilon_2(\omega)$ by Kramers-Kronig transformation [18].

$$\varepsilon_1(\omega) = 1 + \frac{2}{\pi} P \int_0^\infty \frac{\omega' \varepsilon_2(\omega')}{\omega'^2 - \omega^2} d\omega$$

where P is the principle value of the integral.

Using complex dielectric function, absorption coefficient $\alpha(\omega)$, refractive index $n(\omega)$, energy loss spectrum $L(\omega)$ and photoconductivity $\sigma(\omega)$ can be derived [19].

## 3. Results and discussion

### 3.1. Structural analysis

Marcasite and pyrite structures of FeSe$_2$ were fully optimizied with the above parameters described within the DFT calculations. Our calculations predict that marcasite is the most stable phase and the relative energy difference between two phases is 37 meV per formula unit. The calculated cell parameters for marcasite FeSe$_2$ (m-FeSe$_2$) are 4.8250 (4.8002) Å, 5.8499 (5.7823) Å and 3.6649 (3.5824) Å and for pyrite FeSe$_2$ (p-FeSe$_2$) is 5.8633 (5.7830) Å. The difference between experimental values and calculated lattice parameters are below 3% and is acceptable due to the fact that GGA based DFT calculations usually overestimate the lattice parameters [20]. In marcasite FeSe$_2$ (m-FeSe$_2$), among the six Fe-Se bonds that comprising the iron octahedron (Fig. 1a), four bonds in the equatorial plane (Fe-Se$_{eq}$) are smaller (2.4041 Å) compared to the other two bonds (2.4256 Å) along axial direction (Fe-Se$_{ax}$), and calculated bond length of Se-Se dimer is 2.5017 Å. Whereas, in pyrite FeSe$_2$ (p-FeSe$_2$) all Fe-Se bonds are of equal length (2.4408 Å) and Se-Se dimer length is 2.4308 Å.

### 3.2. Band structure and bonding

The calculated band structure of m-FeSe$_2$ along the high symmetry direction in the Brillouin zone is shown in Fig. 2(a). An indirect band gap of 0.86 eV, observed between valence band maximum (VBM) at Y and conduction band minimum (CBM) at U, is lower than experimentally reported value of 1 eV and the direct band gap is 1.29 eV at G point. DFT exchange correlation functionals are not exact, but being approximate, generally underestimate the band gap of semiconductors and insulators [18]. From the total and partial density of states (PDOS) shown in Fig. 3 (a) it is clear that two bands extending from the energy range -16 eV to -11 eV are primarily from Se-4s states and a small contribution from Fe-4s states. Region from -1 eV to -7 eV is characterized by hybridized Fe-3d and Se-4p orbitals. Upper part of valence band is dominated by Fe-3d states but Fe-4s, Se-4p and Se-4s states also exist. Conduction band is mainly from hybridized Fe-3d and Se-4p states, and a little contribution from the 4s states of Fe and Se are present. Band structure and density of states (DOS) of p-FeSe$_2$ are shown in Fig. 2(b) and 3(b) respectively. We observe an indirect band gap of 0.68 eV (X $\rightarrow$ G) for p-FeSe$_2$, lower than that of m-FeSe$_2$ and a direct band gap of 0.71 eV at the G point. The conduction band originates from hybridized Fe-3d and Se-4p states. The valence bands below Fermi energy level upto -6 eV is divided into two bands and are characterized by Fe-4s, Se-4p and Fe-3d states, where the hybridization of Fe-3d and Se-4p states dominates in the upper part. A small contribution of Fe-4s and Se-4s states are also seen along the conduction and valence bands. The separation between two bands (mainly from Se-4s states) in the energy region between -10 eV and -16 eV is increased from 0.37 eV (marcasite) to 1.34 eV (pyrite). Fe-Se bonds in m-FeSe$_2$ are smaller than in p-FeSe$_2$, while Se-Se bonds are longer in m-FeSe$_2$ than in p-FeSe$_2$. Though Fe and Se have same co-ordination in both structures, the differences in bond lengths leads to different band structure.

In order to analyze bonding character of orbitals, bond overlap population (BOP) is calculated based on Mulliken population analysis [21]. BOP can be used as a measure of bonding in solids, where a high (low) value indicates covalent (ionic) bond. The calculated BOP value of 0.11 for Fe-Se$_{eq}$ bonds in the equatorial plane of m-FeSe$_2$ reveals more ionic nature. The Fe-Se$_{ax}$ bonding along the axial line has BOP value of 0.69, suggesting that Fe-Se$_{ax}$ bond has high degree of covalency. The BOP value of Se-Se dimer is equal to 0.59, indicates the mixed covalent and ionic characteristics. For p-FeSe$_2$, the observed BOP value of Fe-Se bond is 0.17 manifests the more ionicity of bonds. Se-Se has a higher BOP value of 1.15, comparable with that of covalent C-C bond in diamond (1.08). This indicates a highly covalent interaction between Se-Se dimer pair in p-FeSe$_2$ due to its shorter bond length than that in m-FeSe$_2$.

Dispersion of energy levels at VBM and CBM can be used to characterize the electronic behaviour of materials. Dispersion of bands at the top of VB and bottom of CB is nearly same for m-FeSe$_2$. Bands are less dispersive at VBM than that at CBM for p-FeSe$_2$ and can indicate n-type semiconducting behaviour. Effective masses (m$^*$) of electrons and holes are calculated at k points by fitting the band curves using the equation [23]

$$m^* = \hbar^2 \left[\frac{d^2E}{dk^2}\right]^{-1}$$

and is listed in Table 1. Electron and hole effective masses are small and rather similar in the case of m-FeSe$_2$. Hole effective masses are larger than electron effective masses in p-FeSe$_2$ suggesting that the material favours an n-type conductivity. Such light mass charge carriers can be easily separated due to photoexcitation and can enhance the efficiency of solar cell devices.

*3.3. Optical properties*

The studies of optical properties for FeSe$_2$ are of interest in the view of potential uses of these materials in solar cells and optoelectronics. For calculating optical properties, we have considered plane polarized light incident on different crystal directions [100], [010] and [001]. The real and imaginary part of dielectric function, $\varepsilon_1(\omega)$ and $\varepsilon_2(\omega)$ as a function of incident photon energy along three crystallographic directions ([100], [010] and [001]) for m-FeSe$_2$ and p-FeSe$_2$ are shown in Fig. 4 and Fig. 5. Both $\varepsilon_1(\omega)$ and $\varepsilon_2(\omega)$ of m-FeSe$_2$ exhibit anisotropy with respect to light polarization direction. The second derivative of real and imaginary part of dielectric function of m-FeSe$_2$ along [100] direction (inset of Fig. 4), has the lowest critical point at around 1.50 eV (A$_m$) which arises from direct transition between highest valence and the second lowest conduction band along G. Similar value has been observed in pyrite-FeS$_2$ [24] and it is essential for applications of materials in solar cells. Other critical points in the lower energy region are at 3 eV (B$_m$) and 4.40 eV (C$_m$), arising from transition between valence and conduction band (indicated by arrows in the band structure). In the higher energy region, the absorption points 6.65 eV ($D_m$), 8.33 eV ($E_m$), 11.22 eV ($F_m$) and 12.90 ($G_m$) represent transitions between Fe-3d and Se-4p states or Se-4p and Fe-4s states. Owing to anisotropy of m-$FeSe_2$, there is a shift in critical points along [010] and [001] polarization direction (not shown). Due to the cubic symmetry of p-$FeSe_2$, the dielectric function is isotropic with respect to different polarization directions. The second derivative of real and imaginary part of dielectric function of p-$FeSe_2$ along three polarization directions are shown as inset of Fig. 5. Lowest critical point around 1.43 eV ($A_p$) is due to the direct transition between highest valence and the second lowest conduction band along G. Other critical points in the lower energy region at 2.93 eV ($B_p$) and 4.41 eV ($C_p$), are due to various transitions from valence band to conduction band (represented by arrows). In the higher energy region, the critical points at 6.54 eV ($D_p$), 8.23 eV ($E_p$), 11.17 eV ($F_p$) and 12.86 ($G_p$) are due to transitions between Fe-3d and Se-4p states or Se-4p and Fe-4s states. It is noted that, in both polymorphs, the optical points donot correspond to a single interband transition since many transitions can be found in the band structure.

Static dielectric constant $\varepsilon_1(0)$ for m-$FeSe_2$ is highest along [100] direction with a value of 22. Along the [010] and [001] polarization directions, it is found to be 20 and 17.5 respectively and is 18.74 for p-$FeSe_2$. Narrow band gap semiconductor such as PbS and $FeS_2$ have showed similar values of static dielectric constants [25,26]. $\varepsilon_1(\omega)$ reaches its maximum value at 1.24 eV for m-$FeSe_2$ and 1.35 eV for p-$FeSe_2$. Dielectric function becomes negative at certain range of energies for both the polymorphs, where the material shows metallic behaviour.

Absorption coefficient $\alpha(\omega)$, refractive index $n(\omega)$, energy loss spectrum $L(\omega)$ and photoconductivity $\sigma(\omega)$ of m-$FeSe_2$ and p-$FeSe_2$ are shown in Fig. 6 and 7 respectively. Optical absorption coefficient is an important parameter for calculating the performance characteristics of solar cells. Absorption spectrum span from 0.86 to 30 eV and from 0.67 to 21 eV for marcasite and pyrite respectively. Both the polymorphs exhibit absorption coefficient ($10^5$ cm$^{-1}$) in the visible region and this value is comparable with other II-VI semiconductors [27]. The energy loss-function, $L(\omega)$, describes the energy loss of a fast electron traversing in the material and the peaks in $L(\omega)$ spectra correspond to plasma oscillations. The strongest peaks of m-FeSe$_2$ in the $L(\omega)$ spectrum are at 24.84 eV, 24.49 eV and 25.02 eV in [100], [010] and [001] directions respectively, which corresponds to Plasmon resonance where $\varepsilon_2(\omega) \ll 1$ and $\varepsilon_1(\omega)$ crosses zero point [28]. The shoulders appearing below this energy originated due to interband transitions [29]. In addition to this, small peaks appear at 6.74 and 6.10 eV along [010] and [001] directions indicate Plasmon resonance at low energy. In the case of p-FeSe$_2$, characteristic Plasmon resonance occurs at a lower energy of 21.53 eV along all the three directions. Another important parameter for design of optical materials is refractive index, which is closely related to the polarizability of ions and local field inside the material. The refractive index along [100], [010] and [001] direction is found to be 4.62, 4.48 and 4.12 in m-FeSe$_2$ and 4.30 in p-FeSe$_2$. Such higher refractive index values are shown by pyrite FeS$_2$ [30]. Photoconductivity is used to define the increase in electrical conductivity of a material, due to the creation of charge carriers as the material absorbs incident photons. An increase in photoconductivity with respect to incident photon is observed when the incident photon energy exceeds the band gap of the material. The material is photoconductive upto 25 eV for m-FeSe$_2$ and upto 20 eV for p-FeSe$_2$, hence these polymorphs can be exploitable for solar cell absorber.

4. **Conclusion**

We have presented a detailed investigation on the electronic structure, bonding nature, optical properties of marcasite and pyrite FeSe$_2$ using first principles calculations. Optimized lattice parameters are in good agreement with experimental results. Electronic density of states show that both m-FeSe$_2$ and p-FeSe$_2$ are indirect band gap semiconductors with band gaps of 0.86 eV and 0.67 eV respectively. Bonding is of mixed ionic and covalent nature in both the polymorphs. Electron and hole effective masses are similar for m-FeSe$_2$. Hole effective masses are larger in p-FeSe$_2$ favouring n-type character. m-FeSe$_2$ exhibits small optical anisotropy with respect to polarization directions whereas p-FeSe$_2$ is optically isotropic. The calculated optical properties of both marcasite-FeSe$_2$ and pyrite- FeSe$_2$ indicate that these materials could be a potential material for applications in solar cells and optoelectronics and it will stimulate further experiments on polymorphs of FeSe$_2$.

**Acknowledgements**

We gratefully acknowledge Department of Biotechnology (DBT), India for the financial support to buy CASTEP. Computational support from High Performance Computing Environment (HPCE), IIT Madras is gratefully acknowledged.

**Table**

Table 1. Effective masses of electrons (conduction band) and holes (valence band) of m-FeSe$_2$ and p-FeSe$_2$ along high symmetry directions. Masses are given in units of free electron mass, m$_o$.

|  |  | Valence Band |  | Conduction Band |
|---|---|---|---|---|
| m-FeSe$_2$ | G-Z | 0.057 | G-Z | 0.12 |
|  | T-Y | 0.11 | T-Y | 0.033 |
|  | Y-S | 0.024 | S-X | 0.035 |
|  | S-X | 0.057 | X-U | 0.036 |
|  | X-U | 0.061 | U-R | 0.036 |
| p-FeSe$_2$ | G-R | 0.11 | G-R | 0.06 |
|  | R-M | 0.23 | G-M | 0.046 |
|  | X-R | 0.14 | R-M | 0.064 |

**Figures**

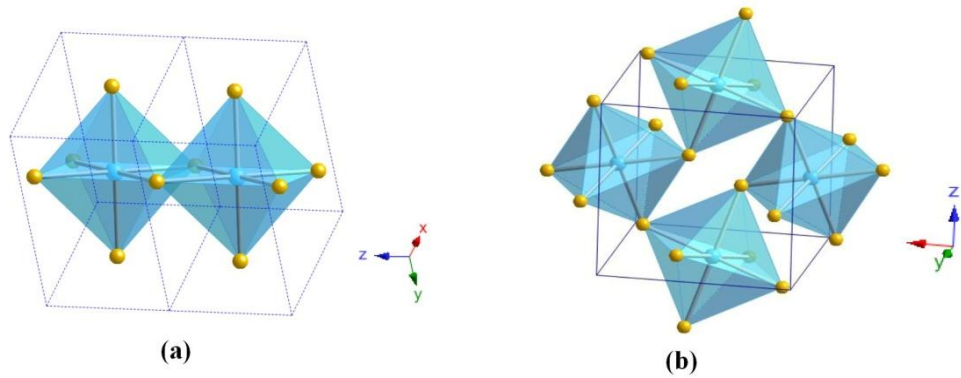

Figure 1: (a) Crystal structure of m-FeSe$_2$. Edge sharing octahedra forming a chain parallel to c-axis of m-FeSe$_2$. (b) crystal structure of p-FeSe$_2$ in which adjacent octahedra are connected by corner. (blue and yellow balls represent Fe and Se atoms respectively).

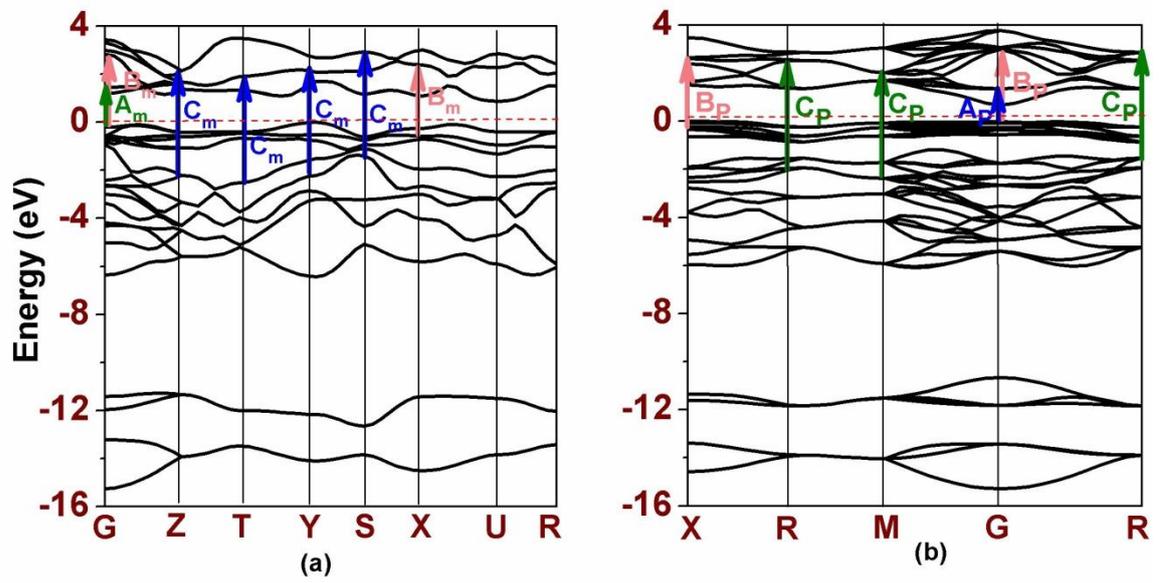

Figure 2: Calculated Band structures of (a) m-FeSe$_2$ and (b) p-FeSe$_2$

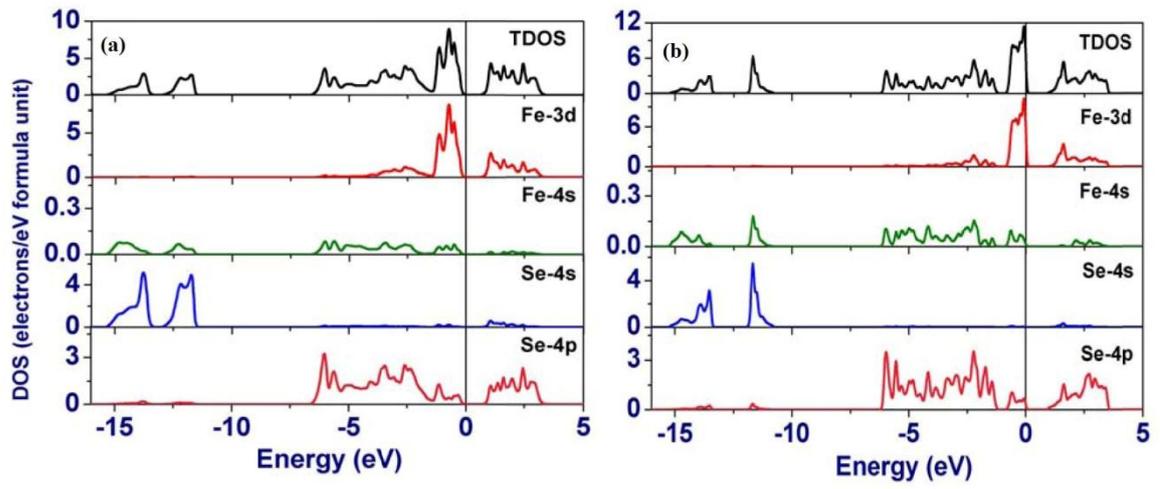

Figure 3: Density of States of (a) m-FeSe$_2$ and (b) p-FeSe$_2$

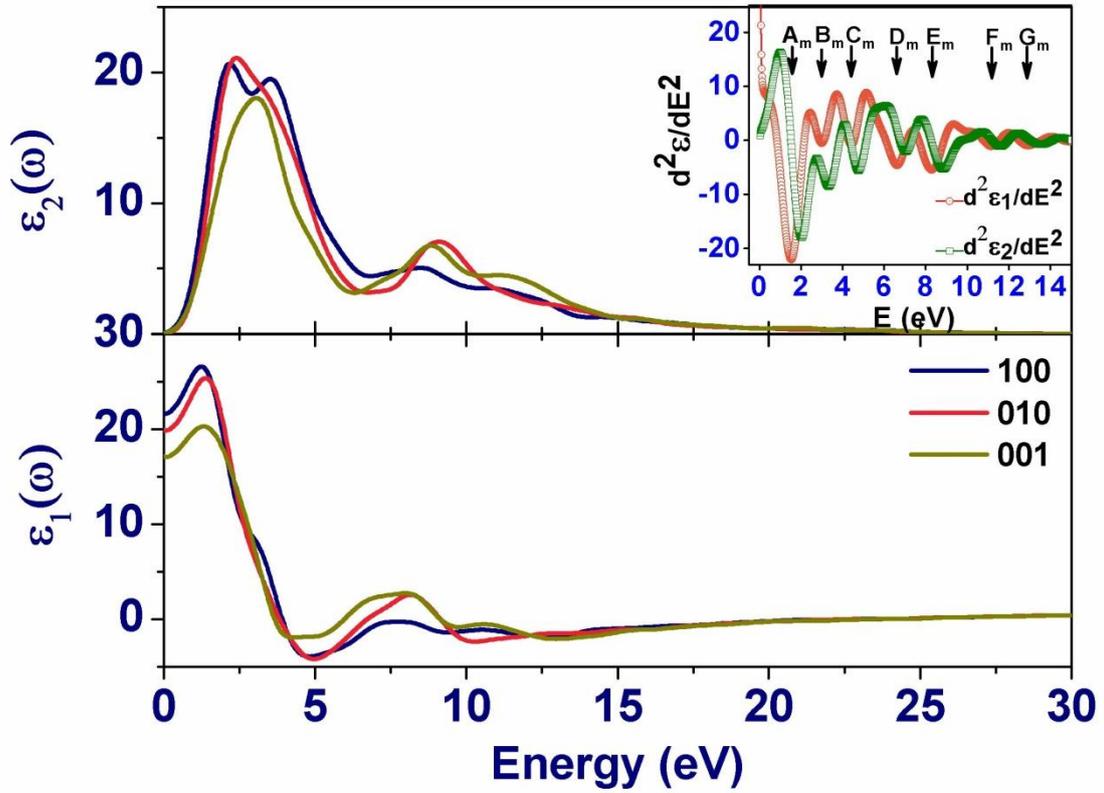

Figure 4: Real and imaginary part of dielectric function $\varepsilon(\omega)$ of marcasite FeSe$_2$ along [100], [010] and [100] direction. Inset shows calculated second derivative of dielectric function of m-FeSe$_2$ with respect to energy.

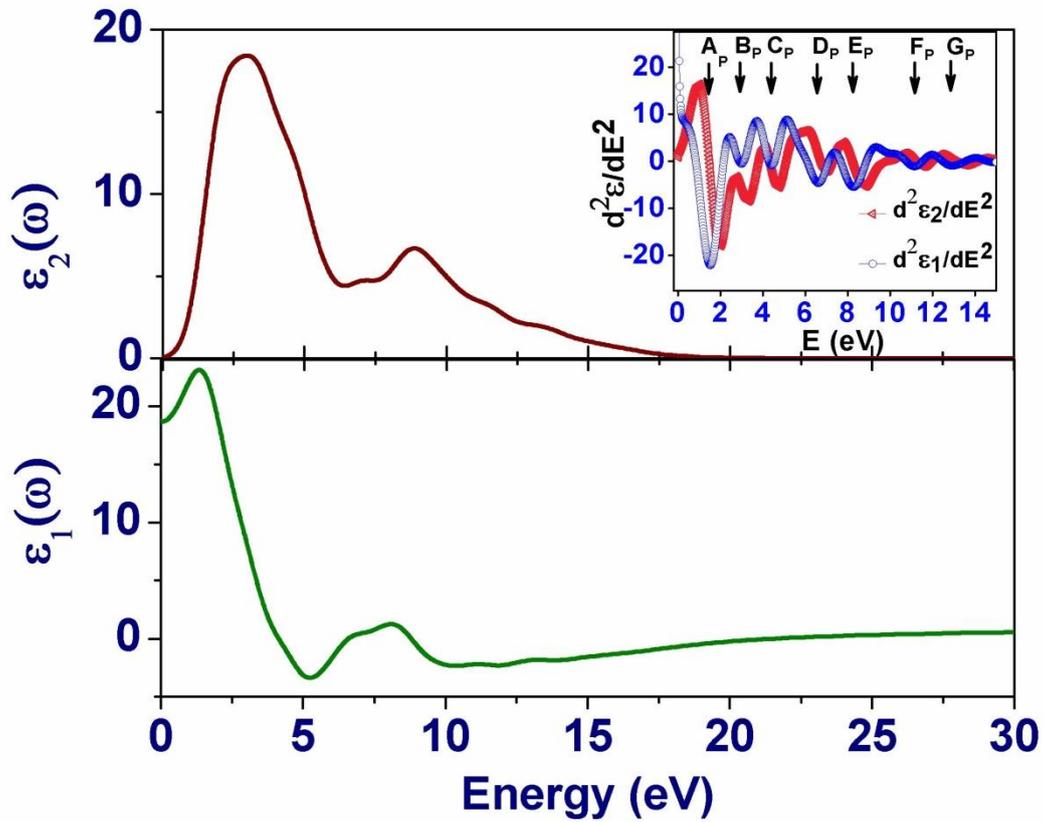

Figure 5: Real and imaginary part of dielectric function $\varepsilon(\omega)$ of pyrite FeSe$_2$ along [100] direction. Properties are similar in [010] and [001] directions also. Inset shows calculated second derivative of dielectric function of p-FeSe$_2$ with respect to energy.

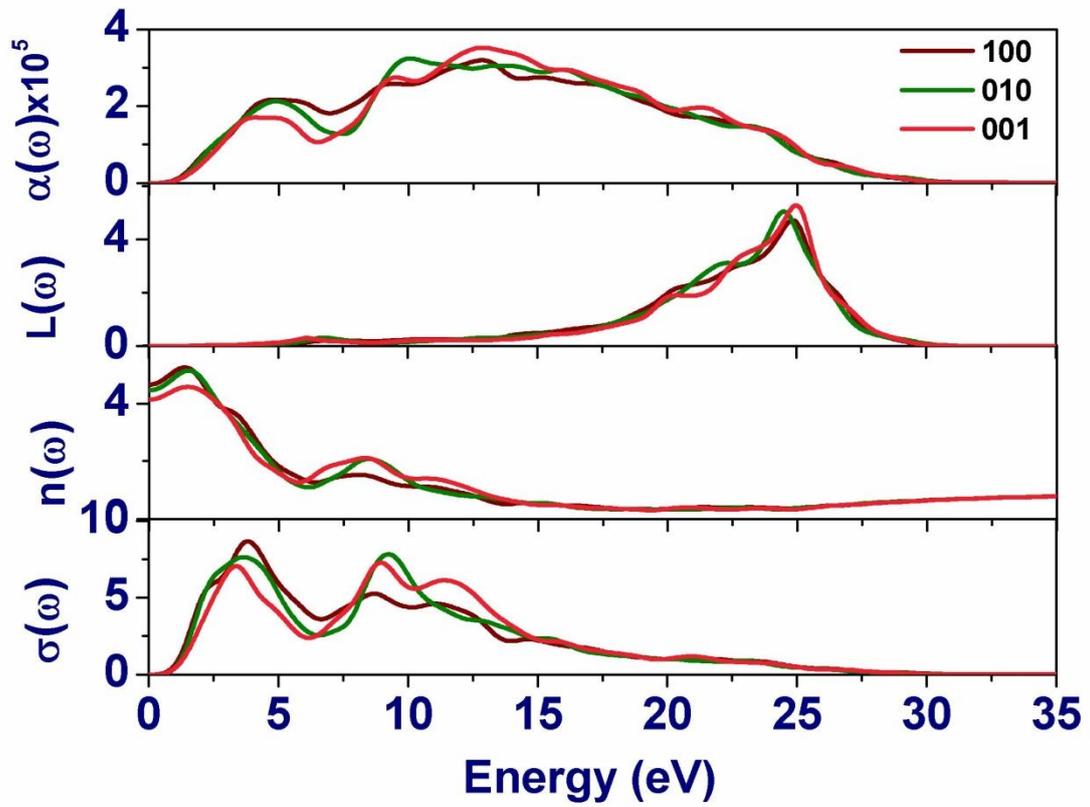

Figure 6: Absorption coefficient $\alpha(\omega)$ energy-loss function $L(\omega)$, refractive index $n(\omega)$ and conductivity $\sigma(\omega)$ of m-FeSe$_2$ along [100], [010] and [100] direction.

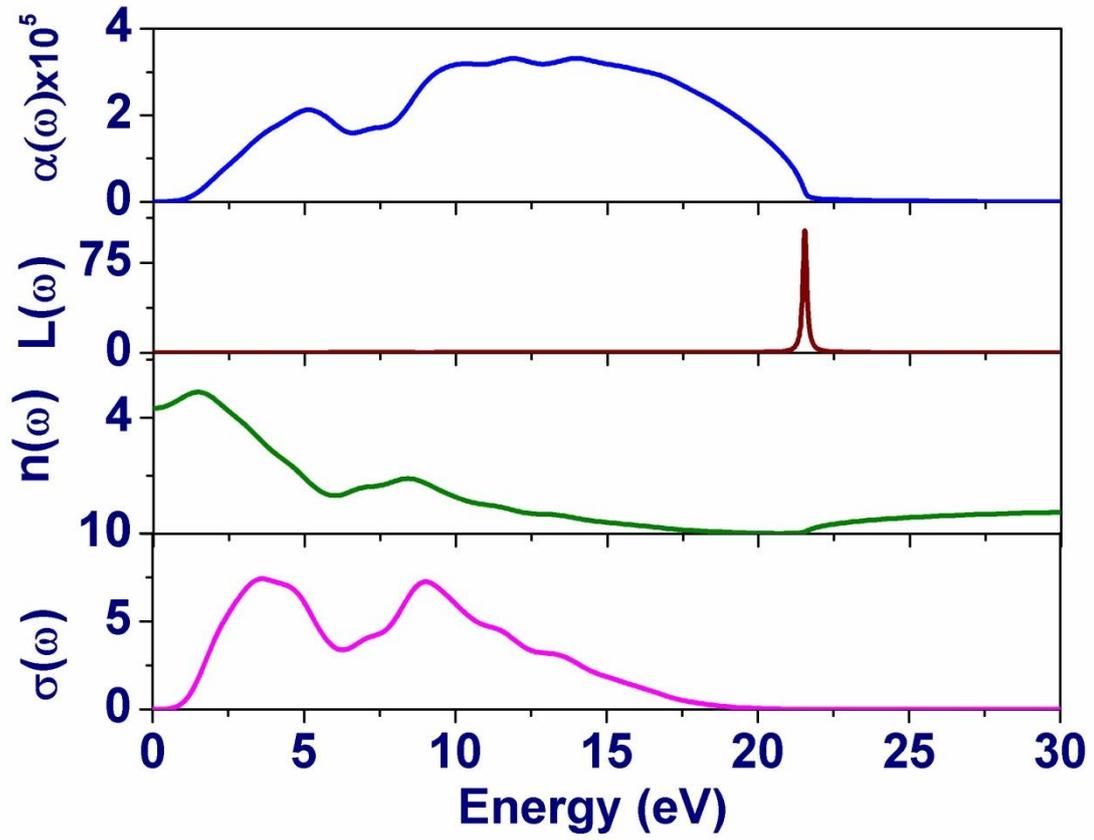

Figure 7: Absorption coefficient $\alpha(\omega)$ energy-loss function $L(\omega)$, refractive index $n(\omega)$ and conductivity $\sigma(\omega)$ of p-FeSe$_2$ along [100] direction. Properties are similar along [010] and [001] directions also.